\documentclass[pra,twocolumn,nofootinbib,superscriptaddress]{revtex4-1}
\usepackage{graphicx}
\usepackage{dcolumn}
\usepackage{bm,bbm}
\usepackage{xcolor}
\usepackage{tcolorbox}
\usepackage{algorithm}
\usepackage{algpseudocode}
\usepackage{amsmath}
\usepackage{bbold}
\usepackage{braket}
\usepackage{nameref}
\usepackage[toc,page]{appendix}

\usepackage[colorlinks,breaklinks,linkcolor={blue},citecolor={magenta},urlcolor={blue}]{hyperref}
\usepackage{enumerate}%allows different styles of enumerate environment

\makeatletter
\newcommand\org@hypertarget{}
\let\org@hypertarget\hypertarget
\renewcommand\hypertarget[2]{%
  \Hy@raisedlink{\org@hypertarget{#1}{}}#2%
  }
\makeatother

\begin{document}
\title{Spin-orbit coupling induced by ascorbic acid crystals}
\author{Florence Grenapin}\affiliation{Department of physics, University of Ottawa, Advanced Research Complex, 25 Templeton Street, K1N 6N5, Ottawa, ON, Canada}
\author{Alessio D'Errico}
\email{ aderrico@uottawa.ca}
\affiliation{Department of physics, University of Ottawa, Advanced Research Complex, 25 Templeton Street, K1N 6N5, Ottawa, ON, Canada} 
\author{Ebrahim Karimi}\affiliation{Department of physics, University of Ottawa, Advanced Research Complex, 25 Templeton Street, K1N 6N5, Ottawa, ON, Canada}
\affiliation{National Research Council of Canada, 100 Sussex Drive, K1A 0R6, Ottawa, ON, Canada}
\begin{abstract}
Some anisotropic materials form semicristalline structures, called spherulites, which observed in a polarisation microscope, exhibit a characteristic ``maltese-cross"-like pattern. While this observation has been hitherto considered as a tool to characterize these materials, we show that these patterns are associated with a strong light's spin-orbit coupling induced by the spherulite structures. We experimentally demonstrate these effects using samples of crystallized ascorbic acid and observing the creation of optical vortices in transmitted laser beams, as well as the formation of inhomogeneous polarisation patterns. Our findings suggest the use of spherulites in frequency ranges, e.g. in the THz domain, where polarisation and spatial shaping of electromagnetic radiation is still a challenging task.
\end{abstract}

\date{June 2022}
\maketitle
\section{Introduction}
\begin{figure*}[htbp]
\centering
\includegraphics[width=\textwidth]{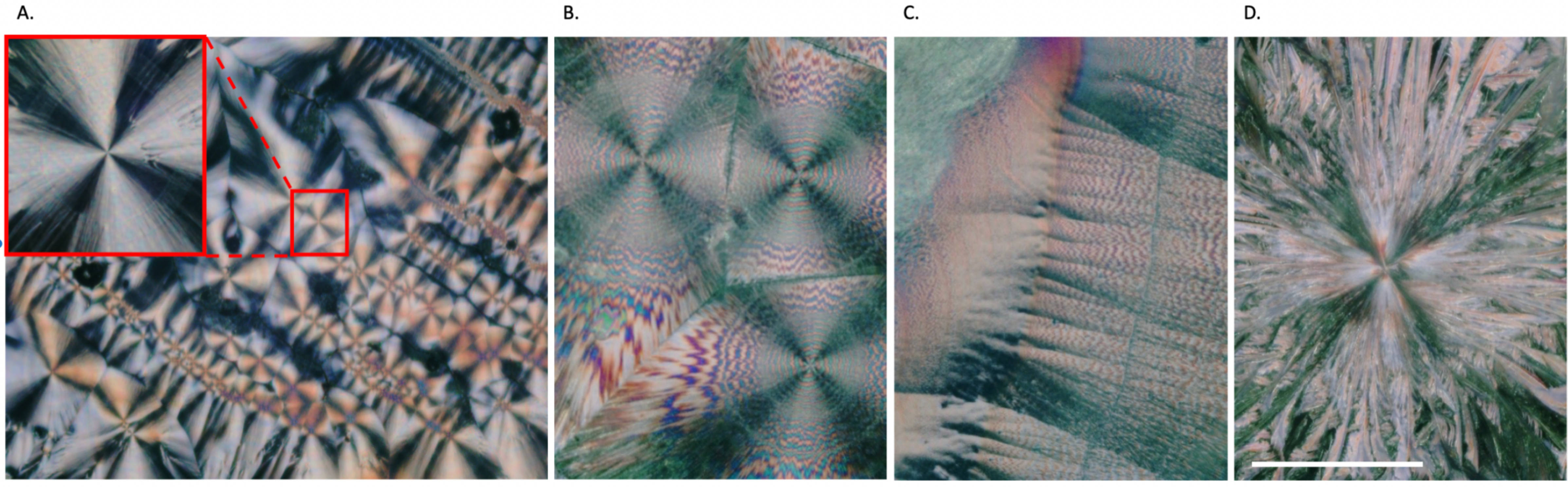}
\caption{Images of ascorbic acid samples between cross polarizers, illuminated by white light. \textbf{A.} Sample of ascorbic acid crystals grown at room temperature, from a 10:1 ethanol:water solution. Many ``maltese-cross"-like patterns are visible, delimited by boundary regions. In the top left corner, a singled out pattern is enlarged. \textbf{B.} Sample of ascorbic acid crystals grown at $80^{\circ}$C, from a pure water solution. The ``maltese-cross"-like patterns exhibit a double-banded structure visible as concentric rings of alternating colors. \textbf{C.} Sample grown in the same conditions as B. Point defects aligned close together on a physical boundary cause a different pattern to arise in the sample. The double-bandedness remains visible, without the cross pattern. \textbf{D.} Sample of ascorbic acid crystals grown at $120^{\circ}$C, from a pure water solution. A "maltese-cross"-like pattern is visible, but appears disorganized and coarse. The fibrils themselves are almost recognizable. The scale bar corresponds to $\sim$0.5 mm.}
\label{fig:1}
\end{figure*}

Polymer spherulites are spherically symmetric semicristalline structures typically observed when a molten polymer is slowly cooled down \cite{crist2016polymer}. The slow cooling allows polymer chains to form in ordered configurations. The crystallization starts around point defects and consists in lamellae structures that, in the absence of temperature gradients, grow radially from the defect centre. 
\\
The radially directed fibrillar structure can be easily seen at low magnification. Such fibrils are composed of one or multiple crystals elongated in the radial direction. As a consequence of the optical anisotropy of the crystals, spherulites can be considered as one of the simplest examples of patterned anisotropic media. Such media have been subject of intense research in the last decade due to their action on optical beams that couples optical polarisation and inhomogeneous phase transformations. More specifically, Pancharatnam-Berry Optical Elements (PBOEs) \cite{marrucci2006pancharatnam} are slabs of patterned anisotropic materials whose inhomogeneous orientation of the optic axis can be used to structure the phase and polarisation of light in a way that is conditioned by the polarisation of the input beam. The most known and widely used device of this kind is the $q$-plate, a PBOE with an optic axis that varies linearly with the azimuthal angle \cite{marrucci:06}. $Q$-plates have been used as a tool in different areas \cite{rubano2019q}: from quantum information and simulation \cite{nagali:09,Cardano2015} to microscopy and surface structuring \cite{yan2015q, nivas2015direct}. Typically $q$-plates are based on liquid crystals, in which case their retardation can be finely controlled either thermally \cite{karimi2009efficient} or electrically \cite{Piccirillo2010}. More recently $q$-plates, as well as other PBOEs, based on subwavelength structures have been developed: from nanostructured glasses \cite{beresna2011radially} to plasmonic metasurfaces \cite{karimi2014generating}.
\\
Here, we show how slabs of spherulites -- specifically how thin layers of ascorbic acid spherulites -- also behave like PBOEs with an azimuthal optic axis dependence. 
Indeed, when observed in the context of polarized light microscopy, i.e. between crossed polarizers, spherulites display ``maltese-cross''-like patterns which strongly resemble $q$-plate patterns with topological charge $q=1$ (see Fig. \ref{fig:1}). \\
In $q$-plates, the radial optic pattern introduces an azimuthal phase factor to the light transmitted through it, a factor which is associated with a well-defined amount of Orbital Angular Momentum (OAM). This process is known as optical spin to orbit coupling. Depending on the input polarisation state one can generate beams with OAM values of $=\pm2$ or induce singular polarisation patterns, e.g. vector vortex beams. OAM beams, as well as vector vortex beams, have been exploited to encode classical and quantum information \cite{nagali:09,dambrosio:12, fickler2012quantum, vallone:14, willner2017recent, Sit:18}.
\\
Spherulite-like structures are observed when evaporating solutions of ascorbic acid dissolved in water (or other solvents, like ethanol). We confirmed our prediction by performing interferometric experiments to ascertain the OAM character of the transmitted beam and spatial Stokes polarimetry to reconstruct the generated polarisation patterns during and after the spherulite formation.

%%%%%%%%%%%%%%%%%%%%%%%%%%%%%%%%%%%%%%%%%%
%%%%%%%%%%%%%%%%%%%%%%%%%%%%%%%%%%%%%%%%%%

\section{Results}
\begin{figure*}[htbp]
\includegraphics[width=2\columnwidth]{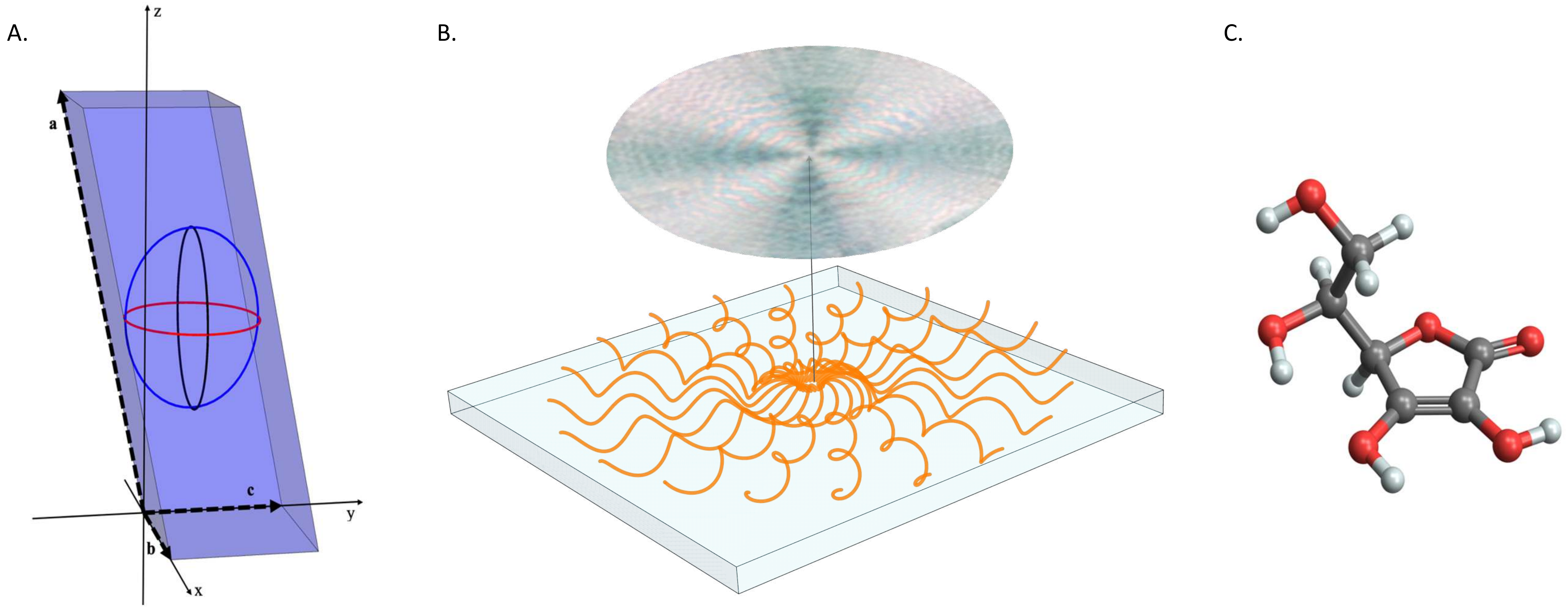}
\caption{\textbf{A.} Schematic of the ascorbic acid crystal unit cell and indicatrix ellipsoid. The three axes of the unit cell are: $a=17.299(8)$ \AA, $ b=6.353(3)$ \AA, $c=6.411(3)$ \AA, with $\omega=102^{\circ}11'(08')$ the angle between $a$ and $c$. The three axes of the ellipsoid are: $\alpha=1.46,\, \beta=1.6,\, \gamma=1.75$, with $\alpha$ parallel to $b$. \textbf{B.} Twisting of indicatrix ellipsoid along the fibril growth directions. \textbf{C.} The molecular structure of ascorbic acid: grey, red and white spheres represent carbon, oxygen and hydrogen atoms respectively.}
\label{fig:2}
\end{figure*}

Ascorbic acid ($C_6H_8O_6$), commonly referred to as vitamin C, forms biaxial crystals with monoclinic spheroidal crystal units of $P2_1$ symmetry.  \cite{cox1932crystalline, hvoslef1968crystal}. The unit cell and indicatrix ellipsoid are shown in Fig. \ref{fig:2}-\textbf{A}, and the molecular structure of ascorbic acid is shown in Fig. \ref{fig:2}-\textbf{C}. The molecules themselves are roughly planar and at room temperature growth they are almost strictly oriented in the (010) plane (the plane generated by the axes $a$ and $c$ in Fig. \ref{fig:2}-\textbf{A}), the smallest refractive index of the indicatrix is aligned in the $b$ direction.\\
In our experiment, commercially available ascorbic acid was spread over a glass slide, dissolved in a solution which was made to evaporate and leave free and dispersed ascorbic acid molecules in the sample. The choices of solvent and temperature are important as they determine the thickness of the sample (which affects the retardation), the rate of crystal formation, subtle variations in crystal structure, and the amount of molecules present on the slide (whether or not the effects will be visible). Water and ascorbic acid have great affinity; however, due to its high surface tension water does not spread out as well on the slide and can take up to a couple hours to dry, at room temperature \cite{uesaka2002pattern}. Therefore at first, we chose to dissolve the ascorbic acid at room temperature in a solution of 10:1 ethanol:water \cite{omar2006effect}. On the sample, the spherulite structures formed around point defects from residual powder grains, either accidentally present from undissolved powder grains in the solution or introduced manually after evaporation. \cite{uesaka2002pattern, ito2003morphological, yamazaki2009humidity}.
There is a large variety of patterns that can be observed under different conditions of humidity and temperature. \cite{yamazaki2009humidity, crist2016polymer}. In particular, in specific humidity conditions (relative humidity $RH\approx 30-40\%$) and at room temperature \cite{yamazaki2009humidity}, one can observe ``maltese-cross"-like patterns when viewing the sample between crossed polarizers and under white light illumination (see Fig. \ref{fig:1}- \textbf{A}). Under different conditions, for example at slightly higher temperatures, the crystals can form with a twist, rotating around the fibrillar axis at a certain angle, resulting in a periodic rotation of the indicatrix ellipsoid along the radial direction \cite{banded1959Price}. This alternation of adjacent regions of relative high and low birefringence forms ``banded maltese-cross"-like patterns, where, when viewed under cross-polarizers, regularly spaced concentric rings of alternating brightness appear overtop of the cross-like intensity pattern. In the case of ascorbic acid, ``double-banded maltese-cross" patterns form because of the biaxial nature of the rotating ellipsoid [Fig. \ref{fig:1}-\textbf{B} and \textbf{C}]. By choosing pure water as a solvent which we heated at near boiling point, we observed the double banded patterns more clearly. In more extreme conditions at higher temperatures, the spherulite patterns become coarse and less organized (see Fig. \ref{fig:1}-\textbf{D}).
\\
In order to understand the action of ascorbic acid crystals on a light beam around point defects we can adopt the model of a wave retarder with an inhomogeneous optic axis orientation. This model works equally for biaxial media once one fixes the propagation direction of the transmitted light. In such a case there are still two (effective) refractive indexes associated with specific orthogonal polarizations, which can be calculated knowing the wavevector of the incident beam and the orientation of the index ellipsoid (see e.g. \cite{saleh2019fundamentals}). In the following we conventionally define as ``optic axis" the axis corresponding to the highest refractive index experienced by the incident light beam. Assuming the sample sufficiently thin and lying in the plane $z=0$, we have an optic axis angle $\theta$ which is a function of the azimuthal angle and radius in cylindrical coordinates: $\theta(\rho,\phi)$. The action of an anisotropic uniaxial medium can be expressed by the following Jones matrix, written in circular polarization basis:
\begin{align}
    U(\delta,\theta)=\cos\left(\frac{\delta}{2}\right)\begin{bmatrix}1& 0\\0&1\end{bmatrix}+i\sin\left(\frac{\delta}{2}\right)\begin{bmatrix}0& e^{-2i\theta}\\e^{2i\theta}&0\end{bmatrix},
\end{align}
where the parameter $\delta$ is the optical retardation of the medium, $\delta=2\pi(n_1-n_2)d/\lambda$, with $n_{1,2}$ the two effective refractive indices, $d$ the sample thickness and $\lambda$ the wavelength. Note that $\delta$ can be position dependent as well, and in particular, in the case of banded ``maltese-cross"-like patterns.\\

%%%%%%%%%%%%%%%%%%%%%%%%%%%%%%%%%%%%%%%%%%
%%%%%%%%%%%%%%%%%%%%%%%%%%%%%%%%%%%%%%%%%%

At half-wave retardation $\delta=\pi$, the effect of such a medium is to invert the handedness of circular polarisation and introduce a phase factor $\exp(\pm i 2\theta(\rho, \phi))$, where the $+$ sign is in the case of input left circular polarisation, and the $-$ sign for input right circular polarisation. In particular, mediums characterized the optic axis angle anisotropy
\begin{equation}
    \theta_q(\rho, \phi)=q\phi+\theta_0,
    \label{eq:Opticaxis}
\end{equation}
where $q$ is the topological charge of the molecular director field and $\theta_0$ an offset angle, introduce azimuthal phase factors of $\exp(\pm i 2\theta_0)\exp(\pm i 2q\phi)$. Beams characterized by $\exp(i\ell \phi)$ factors are known for carrying a well-defined amount of OAM equal to $N\,\hbar\ell$ (where $N$ is the average number of photons). Therefore, the medium described by $U(\pi, \theta_q)$ is converting the input spin angular momentum (SAM) of each input photon into OAM, a process known as SAM-to-OAM conversion (STOC), first investigated with $q$-plates~\cite{marrucci:06}. Moreover, an input beam that is linearly polarized (or in a generic elliptical polarisation state) will be transformed in a non-separable state of polarisation and OAM which manifests in inhomogeneous polarisation patterns \cite{Cardano:12, cardano2013generation, derrico2017a}. 
\\
For radial patterns, which we can expect from the radially directed fibrillar structure of spherulites, one has $q=1$ and $\theta_0=0$. In general, since the optic axis orientation is defined modulo $\pi$, $q$ can be either an integer or a semi-integer number, $q=1/2$ being thus the smallest topological charge that can characterize such defects. It is well known that, in nature, topological defects typically manifest the smallest possible charge (since this correspond to a smaller defect energy~\cite{chaikin_lubensky_1995}). However, in this case there are no constraints on the molecular inclination along the $z$-axis, so semi-integer topological charges can still induce a discontinuity in the pattern under a $2\pi$ rotation, as is typically observed in umbilical defects~\cite{chandrasekhar_1992,Barboza:15}. To test that our samples of ascorbic acid spherulites have a local optic axis around defects of $\theta = \phi$, we observed both the STOC phenomenon and the generation of inhomogeneous polarisation patterns. \\

\begin{figure*}[htbp]
\centering
\includegraphics[width=\textwidth]{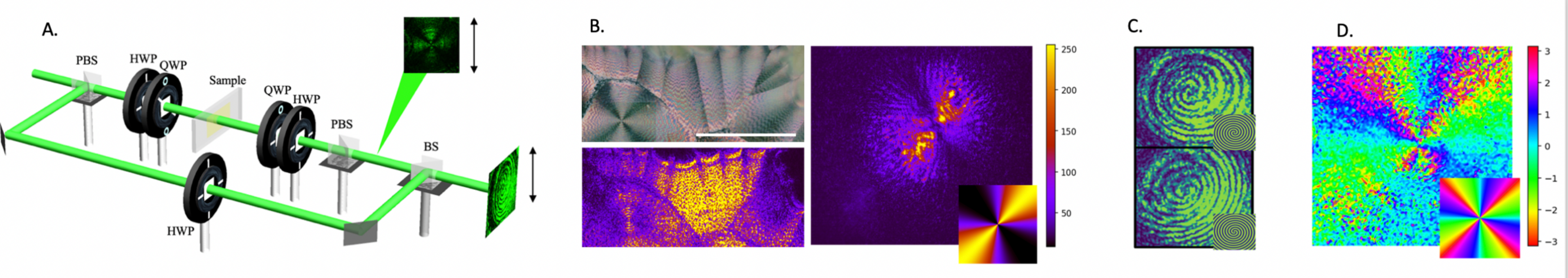}
\caption{\textbf{A.} Schematics of the interferometry and polarimetry setups. The reference beam's polarization is rotated in the reference arm to ensure non-orthogonal polarizations at the output of the BS. A lens (not shown) is also added to the reference arm to induce a spherical wavefront onto the reference gaussian beam. For polarization tomography, the reference beam is ignored and the camera is placed between the PBS and the BS, as shown.
\textbf{B.} Intensity patterns characterizing the optic axis angle in two different regions of the sample. Right-circular polarization is incident on the sample and projected onto horizontal polarization, resulting in an intensity distribution related to the optic axis angle by \ref{eq:theta0meas}. On the right, the distribution around a defect. It matches the simulated intensity for $\theta = \phi$, shown in the bottom right corner. On the bottom left, the intensity distribution of the image above, showing a region near a physical boundary. The intensity does not vary in that region. The scale bar corresponds to ~ 0.5mm.
\textbf{C.} On top, interference pattern generated with input right circular polarization, compared to the theoretical pattern between a spherical gaussian and a beam carrying OAM +2. On the bottom, interference pattern generated with input left circular polarization, compared to the theoretical pattern between a spherical gaussian and a beam carrying OAM -2. \textbf{D.} Density plot of the major axes of the point-by-point polarization ellipses from a beam transmitted through a defect region of the sample. The V-point is visible. A simulation of the expected density plot is shown in the bottom right corner.}
\label{fig:3}
\end{figure*}

\noindent\textbf{Experimental characterization of the optic axis pattern.} We first confirmed that the ascorbic acid crystals orient radially by measuring the optic axis pattern $\theta:=\theta(\rho,\phi)$. This can be done by shining the sample with circular polarisation and projecting the transmitted beam on a linear polarisation state. The resulting intensity distribution is given by 
\begin{equation}
    I(\rho,\phi)\propto 1+\sin(\delta)\sin(2\theta(\rho,\phi)),
    \label{eq:theta0meas}
\end{equation}
where we assumed input right-circular polarisation and a projection on horizontal polarisation. The measurement was performed with a 633 nm diode laser source. At this wavelength our sample exhibits a retardation close to $\delta=\pi/2$, ensuring a good visibility of the intensity modulation in the measured pattern. As shown in Fig. \ref{fig:3}-\textbf{B}, near point defects, the recorded intensity distribution (averaged over different radial positions) is well described by the relation $\theta=\phi$, associated with a radial distribution of the molecular director around the defect. In proximity of boundaries which separate regions containing point defects, we observe a jump in the optic axis angle.\newline
%%%%%%%%%%%%%%%%%%%%%%%%%%%%%%%%%%%%%%%%%%
%%%%%%%%%%%%%%%%%%%%%%%%%%%%%%%%%%%%%%%%%%

\noindent\textbf{STOC in ascorbic acid samples.} A simple way to ascertain the generation of OAM in a laser beam is to look at its interference with a reference Gaussian beam \cite{marrucci:06}. It is easy to show that, in the case of co-propagating beams, the interference pattern consists of spiral fringes where the number of arms is dictated by the OAM absolute value of the analyzed beam, and the handedness by the OAM sign (see insets in Fig. \ref{fig:3}-\textbf{C} for simulations). We thus verified this result by inserting the samples in one arm of an interferometer and recording the interference pattern for right handed and left handed circular polarisation inputs (the experimental setup is shown in Fig. \ref{fig:3}-\textbf{A}). We used a laser beam at 525 nm, where the optical retardation of the samples is typically close to $\pi$. \\
The resulting interference patterns are shown in Fig. \ref{fig:3}-\textbf{C} for right and left circularly polarisations, respectively, showing that the STOC phenomenon is realized.\\

\begin{figure*}[htbp]
\centering
\includegraphics[width=\textwidth]{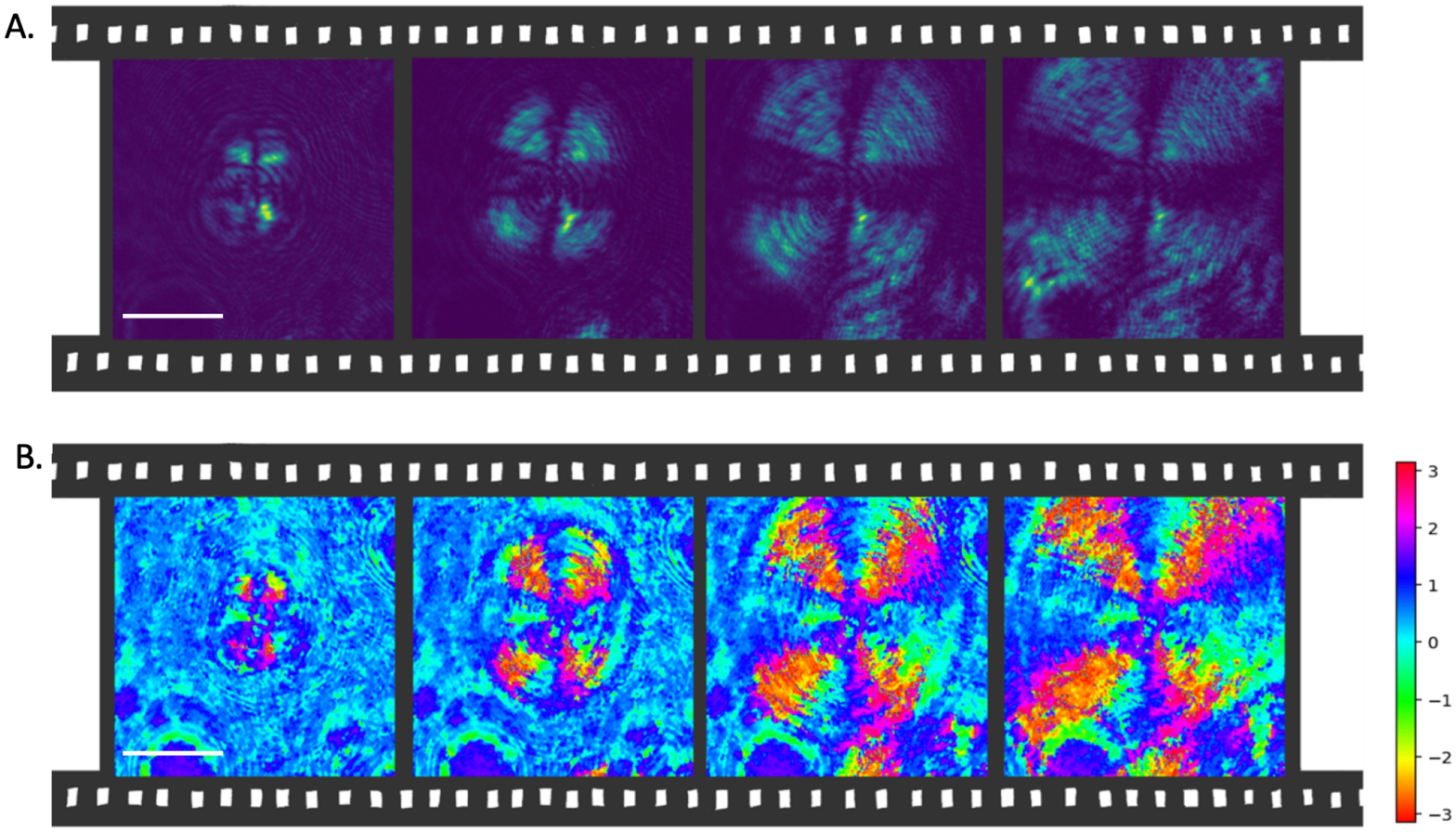}
\caption{Frames showing the evolution of polarization distribution during the growth of the sample at t=0min, 6.5min, 12.8min, 19.1min. The sample was grown at room temperature, from a solution of 10:1 ethanol:water. \textbf{a.} Intensity of the beam between cross polarizers at different times during crystal growth. \textbf{b.} Plot of the major axis angle of the point by point polarization ellipses at different times during crystal growth. The scale bar corresponds to $\sim$1 mm.}
\label{fig:4}
\end{figure*}

\noindent\textbf{Polarimetry of vector modes.} As the next step we illuminated the sample with the same laser beam but with linear polarisation. The $q$-plate model predicts that the resulting field, on the sample plane and at half-wave retardation, is given by
\begin{equation}
    \mathbf{E}_{out}(\rho,\phi)\propto(e^{i2\phi}\mathbf{e}_R+e^{-i2\phi}\mathbf{e}_L)e^{-\rho^2/w^2},
\end{equation}\\
where we assumed a Gaussian envelope field with waist $w$, and $\mathbf{e}_{R,L}=\left(\mathbf{x}\pm i\mathbf{y}\right)/\sqrt{2}$, with $\mathbf{x}$ and $\mathbf{y}$ the unit vectors of the $x$ and $y$ axes. Such a field is linearly polarised everywhere but with an angle of the polarisation ellipse given by $2\phi$ and a singular point on the beam axis (known as a ``V"-point) \cite{cardano2013generation,derrico2017a}. We observed the generation of these patterns by a point-by-point measurement of the Stokes parameters, defined by:
$s_0=\mathcal{I}_{total}$, $s_1=I_H-I_V$, $s_2=I_A-I_D$, $s_3=I_L-I_R$,
where $I_{H,V,A,D,L,R}$ are the intensities of, respectively, horizontal, vertical, anti-diagonal, diagonal, left circular and right circular polarisation, and $\mathcal{I}_{total}$ stands for total intensity. These intensities, and consequently the Stocks parameters, can all be easily measured by a sequence containing a quarter-wave plate, a half-wave plate and a polarizing beam splitter (Fig. \ref{fig:3}-\textbf{A}). By recording the different projection intensities on a camera, we reconstructed the local polarisation ellipse. The resulting pattern, is qualitatively close to the expected one. In particular the density plot of the polarisation ellipse orientation (which is the phase of the field $s_1+i s_2$) clearly shows a singular pattern (Fig. \ref{fig:3}-\textbf{D}). Due to the optical retardation not being exactly $\pi$, the ``V"-point is split in 4 points of circular polarisation (``C"-points) each with topological charge 1/2 (see also ref.~\cite{derrico2017a}). \\
In Fig. \ref{fig:4}, we show some frames of a recording of the Stokes field evolution during the evaporation of the solvent, at room temperature. It can be seen how inhomogeneous polarisation distributions start to develop around the defect and to expand radially. Again, a V-point formation in the polarisation pattern is clearly visible.

%%%%%%%%%%%%%%%%%%%%%%%%%%%%%%%%%%%%%%%%%%
%%%%%%%%%%%%%%%%%%%%%%%%%%%%%%%%%%%%%%%%%%

\section{Discussion and Conclusions}
We have shown that ascorbic acid spherulites crystallize around defects, in the absence of twisting, with a local optic axis pattern of $\theta = \phi$. We have also shown experimentally the generation of OAM modes and vector beams. Our findings show how ascorbic acid can be exploited as a material for fabricating devices which exploit photonic spin-orbit phenomena.\\ 
The radial growth around the defect is likely the most energetically efficient way to form the hydrogen bonds between atoms inside the crystal. It can be interesting to explore the different crystallization structures when ascorbic acid is grown on a polar substrate, instead of around a defect on a neutral glass substrate.\\
Ascorbic acid is part of a large family of spherulites, each of which having structures suggesting certain anisotropy of optic axis patterns around defects. It would be interesting to explore the STOC phenomena in other kinds of spherulites, which may have advantages over ascorbic acid. In particular, recently the controlled generation of quartz spherulites \cite{zhou2021crystallization} may offer an interesting way for creating vortex beams in the Tera Hertz frequency domain \cite{wang2020recent} ($q$plates for this regime have been recently realized using either 3D printing techniques \cite{hernandez2017q} or quartz segmented waveplates \cite{minasyan2017geometric}) It would also shed light on the role of different crystal structures and symmetries in the shaping of light. 
%The expression for the local optical axis orientation of a $q$-plate is given by $\theta(r,\phi)=q\phi+\theta_{0}$. The value of $\theta_0=0$ depends on the orientation of the slow and fast axes with respect to a single molecule of ascorbic acid. Generally speaking, light travelling through double bonds will acquire more retardation than travelling through single bonds; we can therefore expect to see a higher refractive index along the length of the molecule.\\

\section{Acknowledgement}
This work was supported by the Canada Research Chairs (CRC) and Canada First Research Excellence Fund (CFREF) Program, and Joint Centre for Extreme Photonics (JCEP). 

%We have thus shown experimentally the generation of OAM modes and vector beams. Our findings show how ascorbic acid can be exploited as a material for fabricating devices which exploit photonic spin-orbit phenomena. In future we plan to develop techniques to better control the growth of the ascorbic acid samples in ways analogous to what is done with liquid crystals \cite{rubano2019q}.

%%%%%%%%%% If using BibTeX:
%\bibliography{sample}
\bibliographystyle{apsrev4-1fixed_with_article_titles_full_names_new}
\bibliography{main}

\end{document}